\documentclass[11pt]{article}
\usepackage{amsmath,amsfonts,bbm,graphicx,bm}

\newcommand{\ket}[1]{|#1\rangle}
\newcommand{\bra}[1]{\langle#1|}
\newcommand{\mc}[1]{\mathcal{#1}}
\newcommand{\openone}{\mathbbm{1}}
\newcommand{\cs}{\mathsf{C}^*}
\newcommand{\tr}{\mathrm{tr}}

\newcommand{\poly}{\mathrm{poly}}

\begin{document}

\title{\textbf{Complexity of commuting Hamiltonians on a square lattice of
qubits}\vspace*{2ex}}
\author{Norbert Schuch\\[2ex]
\small Institute for Quantum Information\\
\small California Institute of
Technology\\ 
\small MC 305-16, Pasadena CA 91125, U.S.A.}
\date{}

\maketitle 

\begin{abstract}
We consider the computational complexity of Hamiltonians which are sums of
commuting terms acting on plaquettes in a square lattice of qubits, and we 
show that deciding whether the ground state minimizes the energy 
of each local term individually is in the complexity class \textsf{NP}.
That is, if the ground states has this property, 
this can be proven using a classical certificate which can be
efficiently verified on a classical computer.   Different to previous
results on commuting Hamiltonians, our certificate proves the existence of
such a state without giving instructions on how to prepare it.
\end{abstract}

\section{Introduction}

Understanding the ground state properties of spin systems on a lattice is
of central importance in many-body physics, but at the same time, 
it is a highly challenging problem in many scenarios.   An important step in
understanding its difficulty has been the insight that
computing e.g.\ the ground state energy of a classical spin system is, in
general, an \textsf{NP}--complete problem~\cite{barahona:spinglass-np}:
While the energy of any given spin configuration can be easily computed,
finding the configuration with minimal energy is in general a difficult
task -- it can be as hard as any problem in \textsf{NP}, i.e., any problem
whose solution can be efficiently verified. For quantum spin
systems, an additional difficulty arises: Generally, we cannot even expect
to have an efficient description of the ground state. Thus, it seems that
the only statement we can make about the difficulty of the problem is that
given a quantum register with the ground state, we will be able to
efficiently estimate its energy using a quantum computer.  Indeed, it has
been shown that this is the best we can say, as the problem of estimating
the ground state energy of a quantum system is a complete problem for the
class \textsf{QMA}~\cite{kitaev:book,kitaev:qma}, the quantum analogue of
\textsf{NP}: It contains all problems which have a quantum solution which
can be efficiently checked on a quantum computer, and thus, determining
the ground state energy of a quantum spin system is as hard as any of
these problems; in fact, the problem retains its hardness even when 
restricted to two-dimensional lattices
of qubits with nearest-neighbor interactions~\cite{terhal:lh-2d-qma} or
one-dimensional chains~\cite{aharonov:1d-qma}.

It is an interesting question to understand the reasons underlying the
additional complexity of quantum spin systems as compared to classical
systems. To this end, restricted version of the problem which lie between
classical and general quantum spin Hamiltonians have been studied:
For instance, it has been shown that so-called \emph{stoquastic}
Hamiltonians, where all off-diagonal elements are negative, have a
complexity which lies in between \textsf{NP} and \textsf{QMA}, as those
systems can be related to classical random
processes~\cite{bravyi:stoquastic-ham,bravyi:stoq-ffree};  in fact, these
are exactly the Hamiltonians which allow for Quantum Monte Carlo
simulations as they do not exhibit a sign problem.

Another restricted class of Hamiltonians are \emph{commuting
Hamiltonians}, that is, Hamiltonians which can be written as a sum of
mutually commuting few-body terms. For those systems, all terms can be
simultaneously diagonalized, just as for classical systems; however, the
corresponding eigenbasis can be highly entangled, making it unclear
whether a useful classical description of the ground state can be
provided. In fact, commuting Hamiltonians 
encompass systems which exhibit rich non-classical physics, in particular
models with topological order and even
anyonic excitations, such as Kitaev's toric code and quantum double
models~\cite{kitaev:toriccode}, or Levin and Wen's string net
models~\cite{levin-wen:string-net-models}.  Commuting Hamiltonians are
also of interest since the fixed points of
renormalization flows in gapped phases are expected to be commuting
Hamiltonians, and thus understanding their structure might give insight
into the structure of gapped quantum phases. Finally, understanding the
complexity of commuting Hamiltonians is of interest in quantum complexity,
as it might be a step towards a quantum PCP theorem, which would assess
how the difficulty of estimating the ground state energy is related to the
desired accuracy  which is integer for commuting projectors.

What is know about the complexity of finding the ground state energy of
commuting Hamiltonians, or rather, of determining whether the ground state
minimizes each term in the Hamiltonian individually -- the
\textsc{commuting hamiltonian} problem?
For lattices in two and higher dimensions, \textsc{commuting hamiltonian} 
is an \textsf{NP}--hard
problem, as it e.g.\ encompasses the Ising model on a planar
graph~\cite{barahona:spinglass-np}. On the other hand, it is not clear
whether the general \textsc{commuting hamiltonian} problem is
\emph{inside} \textsf{NP}, since it is not clear in general how to provide
an efficiently checkable description of the ground state.
For two-local (i.e., two-body) Hamiltonians, Bravyi and
Vyalyi~\cite{bravyi:comm-ham} have shown that the problem is in
\textsf{NP} by using $\cs$--algebraic techniques (their result also
implies that one-dimensional commuting Hamiltonians are efficiently
solvable); subsequently, Aharonov and
Eldar~\cite{aharonov:commuting-ham-triangles} have proven containment in
\textsf{NP} for Hamiltonians with three-body interactions both for qubits
on arbitrary graphs, and qutrits on nearly-Euclidean interaction graphs.
In both of these works, the classical
certificates do not only prove the problem to be in \textsf{NP}, but can
in fact be used to construct constant depth quantum circuits which
\emph{prepare} the ground state. This, in particular, implies that the
corresponding Hamiltonians -- including qutrits with three-body
interactions -- cannot exhibit topological
order~\cite{aharonov:commuting-ham-triangles,bravyi:lr-bound-topo-order}.
On the other hand, Kitaev's toric code, which is the ground state of a
commuting Hamiltonian with four-body interactions of qubits, does have
topological order, and thus, we cannot expect any approach which yields a
low-depth circuit to work beyond three qutrits.

In this paper, we study the \textsc{commuting hamiltonian} problem on a
square lattice of qubits with plaquette-wise interactions, and prove that
it is in \textsf{NP}. That is, we consider a square lattice of qubits,
with a Hamiltonian with mutually commuting terms acting on the four qubits
adjacent to each plaquette, and show that the problem of deciding whether
its ground state minimizes the energy of each local term in the
Hamiltonian is in \textsf{NP}: i.e., in case the ground state has this
property, a classical certificate exists which can be checked efficiently
by a quantum computer. Our approach differs considerably from the
aforementioned approaches in that the certificate cannot be used to devise
a quantum circuit for preparing the ground state, and is thus also
applicable to systems with topological order; it should be noted that the
same holds true for the proof in Ref.~\cite{bravyi:comm-ham} that
\textsc{commuting hamiltonian} with factorizing projectors is in
\textsf{NP}.

\section{The setup}

We will consider a 2D square lattice with spins on the vertices, and
either open or periodic boundary conditions. The
Hamiltonian 
\[
H=\sum_p h_p
\]
consists of plaquette terms $h_p$ which act on the four spins adjacent to
the plaquette $p$, and we impose that all the terms in the Hamiltonian
commute, $[h_p,h_q]=0\ \forall\,p,q$.  

As the Hamiltonian terms commute, there is a basis of eigenstates of $H$
which are simultaneously eigenstates of all the $h_p$.  We would like to
know the computational difficulty of the following problem, called
\textsc{commuting hamiltonian}: Is there an eigenstate $\ket{\psi}$ of $H$
which minimizes the energies for all $h_p$ individually, i.e., are the
ground states of $H$ also ground states of each $h_p$?  In the following,
we will show that in the case of qubits, the existence of such a state can
be proven within \textsf{NP}, i.e., there is a classical certificate which
proves the existence of such a $\ket\psi$, and which can be checked
efficiently classically.  Note that on the other hand, it is clear that
the problem is \textsf{NP}--hard, as it e.g.\ encompasses classical Ising
spin glasses in a field which are known to be
\textsf{NP}--hard even for two-level systems~\cite{barahona:spinglass-np}.

For the following, it will be useful to reformulate \textsc{commuting
hamiltonian} as follows: Define the \emph{local ground state projectors}
$\Pi_p$ as the projectors onto the ground state subspace of $h_p$;
the $\Pi_p$ commute again, $[\Pi_p,\Pi_q]=0$. Then,
\[
\Pi_{\mathrm{GS}}=\prod_{p\in \mc P}\Pi_p
\]
is the projector onto the subspace spanned by the states which are ground
states of all $h_p$.  Since \textsc{commuting hamiltonian} asks whether
such states exist, it is equivalent to asking whether $\Pi_\mathrm{GS} \ne
0$.

\section{Commuting Hamiltonian in \textsf{NP}}

\subsection{Two layers}

We start by coloring the plaquettes of the square lattice black and white
in a checkerboard pattern, and denote the set of black and white
plaquettes by $\mc P_B$ and $\mc P_W$, respectively.  Let 
\[
\Pi_B=\prod_{p\in \mc P_B} \Pi_p
    \mbox{\quad and\quad}
\Pi_W=\prod_{p\in \mc P_W} \Pi_p
\]
be the projectors onto the joint ground state space of the black
and white $h_p$, respectively; then, \textsc{commuting hamiltonian} 
corresponds to determining whether $\Pi_B\Pi_W\ne 0$.  This is equivalent
to asking whether
\begin{equation}
\label{eq:commham-trace-formulation}
\tr[\Pi_B\Pi_W]\ne 0
\end{equation}
(this can be seen using eigendecompositions of $\Pi_W$ and $\Pi_B$), and
we will consider this formulation of the problem from now on; to prove
\textsc{commuting hamiltonian} is contained in \textsf{NP}, we therefore
need to show that a classical certificate for the validity of 
(\ref{eq:commham-trace-formulation}) can be provided.

A helpful example to keep in mind in the first part of our discussion is
Kitaev's toric code~\cite{kitaev:toriccode}: There,
$\Pi_p=\tfrac12(\openone+Z^{\otimes 4})$ for $p\in \mc P_B$, and
$\Pi_p=\tfrac12(\openone+X^{\otimes 4})$ for $p\in \mc P_W$, with $X$ and
$Z$ the Pauli matrices.

\subsection{The structure of one layer}

In the following, let us study the structure of each layer individually
(we will w.l.o.g.\ choose black). To this end, we will use a result of
Bravyi and Vyalyi based on $\cs$-algebraic
techniques~\cite{bravyi:comm-ham};  a detailed explanation of those
techniques can also be found in~\cite{aharonov:commuting-ham-triangles}.
The basic insight from Ref.~\cite{bravyi:comm-ham} is the structure of two
commuting projectors. Consider two projectors $L\equiv L_{AB}\otimes
\openone_C$ and $R\equiv \openone_A\otimes R_{BC}$
acting on a space $\mc H_A\otimes \mc H_B\otimes \mc H_C$; the two
operators overlap on $\mc H_B$. Now consider the Schmidt decompositions
\[
L = \sum A_L^i\otimes B_L^i \quad \mbox{and} \quad
R = \sum B_R^i\otimes C_R^i\ ,
\]
i.e.\ $\tr[A_L^i(A_L^j)^\dagger]=0$ for
$i\ne j$, and similarly for $B_L^i$, $B_R^i$, and $C_R^i$. Then, $[L,R]=0$
implies that $[B_L^i,B_R^j]=0$ for all $i,j$, and thus, $B_L^i$ and
$B_R^i$ span commuting $\cs$-algebras, cf.~\cite{bravyi:comm-ham}.
Using the standard form of finite $\cs$-algebras, it follows that the
space $\mc H_B$ has a canonical decomposition
\begin{equation}
        \label{eq:cstardec}
\mc H_B = \bigoplus_\alpha \underbrace{\mc H_L^\alpha
    \otimes \mc H_R^\alpha \otimes \mathcal H_Z^\alpha
    }_{=:\mc H_B^\alpha}
\end{equation}
where the $B_L^i$ span the full matrix algebra on $\mc H_L^\alpha$ while
acting trivially on the rest, and correspondingly for $B_R^i$ and $\mc
H^\alpha_R$. 

This shows that the space $\mc H_B$ can be cut into direct sum ``slices''
(the $\alpha$) such that in each slice, $L$ and $R$ act on independent
subsystems.  More formally, there exists a decomposition
$\openone=\sum_\alpha\pi_\alpha$ of $\mathcal H_B$, with $\pi_\alpha$ the
projectors onto $\mc H_B^\alpha$, such that
\[
[\pi_\alpha,L]=0 \mbox{ and }
[\pi_\alpha,R]=0
\]
and thus
\[
L=
\sum_{\alpha,\beta}\pi_\alpha L \pi_\beta=
\sum_\alpha \underbrace{\pi_\alpha L \pi_\alpha}_{=:L^\alpha}
\ ,\quad
R=\sum_\alpha \underbrace{\pi_\alpha R \pi_\alpha}_{=:R^\alpha}\ ,
\]
where $L^\alpha$ and $R^\alpha$ act on different subsystems $\mc
H_L^\alpha$ and $H_R^\alpha$, i.e., factorize. Note that the above
decomposition allows to compute the $\pi_\alpha$ and thus the $L_\alpha$
and $R_\alpha$ efficiently.

Each vertex in the black sublattice is acted upon by exactly two commuting
projectors $\Pi_p$; thus, we can apply the preceding argument to all
vertices to find
decompositions $\pi_{\alpha_v}^v$, $\sum_{\alpha_v}\pi_{\alpha_v}^v=
\openone$, of the Hilbert space at each vertex $v$, such that $\Pi_B$
projected onto the slice 
$\vec\alpha=(\alpha_v)_{v\in V}$ factorizes,
\[
\bigotimes_{p\in\mc P_B}\Pi_p^{\vec\alpha} = 
 \prod_{v\in V} \pi_{\alpha_v}^v\, \Pi_B\, \pi_{\alpha_v}^v 
\]
(this implicitly defines the $\Pi_p^{\vec\alpha}$),
and $\Pi_B$ can be written as 
\begin{equation}
        \label{eq:black-decomposition}
\Pi_B = \bigoplus_{\vec\alpha}
    \bigotimes_{p\in\mc P_B}\Pi_p^{\vec\alpha}
    \equiv \sum_{\vec\alpha}\bigotimes_{p\in \mc P_B}
    \Pi_p^{\vec\alpha}\ .
\end{equation}
Note that in the sum on the right hand side, we implicitly regard
the tensor products as being canonically embedded into the full Hilbert
space, and we will use this convention in the following.

The analogous decomposition can be performed for the white sublattice, yielding
a (in general different!) decomposition
\begin{equation}
        \label{eq:white-decomposition}
\Pi_W = \bigoplus_{\vec\beta}
\bigotimes_{p\in\mc P_W}\Pi_p^{\vec\beta}
    \equiv \sum_{\vec\beta}\bigotimes_{p\in \mc P_W}
    \Pi_p^{\vec\beta}\ .
\end{equation}
Note that in order to distinguish the decomposition for the the black and
the white layer, we strictly use labels $\vec\alpha$ and $\vec\beta$,
respectively; moreover, we denote the projectors decomposing the white
layer by $\bar\pi_{\beta_v}^v$.

E.g., for Kitaev's toric code the $\pi^v_{\alpha_v}$ are projectors onto
the $Z$ eigenstates, and the $\bar\pi^v_{\beta_v}$ onto the $X$ eigenstates.

\subsection{Combining the layers}

Using the stucture of $\Pi_B$ and $\Pi_W$,
Eqs.~(\ref{eq:black-decomposition}) and (\ref{eq:white-decomposition}), we
can rewrite the \textsc{commuting hamiltonian} problem,
Eq.~(\ref{eq:commham-trace-formulation}), as 
\begin{equation}
\label{eq:sum_alphabeta_omega}
0\ne \sum_{\vec\alpha,\vec\beta}
\,\tr\Bigg[
    \,\,\Big(\!\!\bigotimes_{p\in \mc P_B} \Pi_p^{\vec\alpha}\,\Big)
    \Big(\!\!\bigotimes_{p\in \mc P_W} \Pi_p^{\vec\beta}\,\Big)
\Bigg]\ ;
\end{equation}
recall that we regard the tensor products as being canonically embedded
into the full Hilbert space.
Since each of the traces is positive (as it is the trace of the
product of two positive operators), the above is equivalent to
the existence of $\vec\alpha$ and $\vec\beta$ such that $
\Omega(\vec\alpha,\vec\beta)\ne 0$, where
\begin{equation}
\label{eq:def-omega}
\Omega(\vec\alpha,\vec\beta):=\tr\Bigg[
    \,\,\Big(\!\!\bigotimes_{p\in \mc P_B} \Pi_p^{\vec\alpha}\,\Big)
    \Big(\!\!\bigotimes_{p\in \mc P_W} \Pi_p^{\vec\beta}\,\Big)
    \Bigg]\ .
\end{equation}
Thus, we can ask the prover to provide us as a certificate with some
$\vec\alpha$ and $\vec\beta$ for which $\Omega(\vec\alpha,\vec\beta)\ne0$;
if we can moreover show that $\Omega(\vec\alpha,\vec\beta)$ can be
computed efficiently (or rather in \textsf{NP}), this will prove that
\textsc{commuting hamiltonian} is in \textsf{NP}.

Note that $\Omega(\vec\alpha,\vec\beta)$ can be interpreted as the overlap
of the (unnormalized) states $\bigotimes \Pi^{\vec\alpha}_p$ and $\bigotimes
\Pi^{\vec\beta}_p$, both of which are tensor products of states supported
on individual plaquettes, but with different partitions in the two layers.
Computing such an overlap can in general be as hard as contracting
Projected Entangled Pair States (PEPS)~\cite{verstraete:mbc-peps}, i.e.,
\textsf{PP}-hard~\cite{schuch:cplx-of-PEPS}: any PEPS can be expressed
this way by using one layer for the bonds and the other for the PEPS
projections. Of course, the fact that these states arise from two
commuting layers $\Pi_B$ and $\Pi_W$ yields additional constraints, and we
will show in the following that those constraints allow for the efficient
evaluation of $\Omega(\vec\alpha,\vec\beta)$ in the case of qubits.

For Kitaev's toric code, e.g., we could choose
$\vec\alpha=\vec\beta=(0,\dots,0)$: This yields
\[
\Omega(\vec\alpha,\vec\beta)=\tr\big[(\ket0\bra0)^{\otimes N}
(\ket+\bra+)^{\otimes N}\big] = 2^{-N}\ne 0\ ,
\]
proving the existence of a zero-energy ground state; note that this
certificate does not carry any information on how to prepare the state.

\subsection{Computing the overlap}

Let us now show that for a lattice of qubits, the overlap $\mc
O(\vec\alpha,\vec\beta)$ can be computed efficiently. To this end, we will
show that the computation of the overlap can be decomposed into a product
of overlaps of one-dimensional structures which can be computed efficiently.

Let us first consider the black layer.  For each vertex $v$, the
decomposition (\ref{eq:cstardec}) of the local Hilbert space can either be
trivial (no direct sum) or non-trivial, $\openone=\sum\pi_{\alpha_v}^v$. In
the former case, this implies that at most one of the adjacent plaquette
terms $\Pi_p$ acts non-trivially of vertex $v$; in the latter case, the
sum consists of exactly two \emph{one-dimensional} projectors
$\pi^v_{\alpha_v}$, making use of the fact that the Hilbert space at each
site is a qubit, i.e., two-dimensional.  We will denote the set of
vertices with a non-trivial decomposition in the black layer by $\mc F_B$,
and in the white layer by $\mc F_W$ (in which the one-dimensional
projectors are labelled $\bar\pi^v_{\beta_v}$). 

In $\Omega(\vec\alpha,\vec\beta)$, all vertices in $\mc F_B\cup \mc F_W$ 
contribute only a one-dimensional subspace and thus can be traced out: 
That is, all vertices in $\mc F_B\cap \mc F_W$ can be removed (taking care
whether the overlap of the one-dimensional projectors is non-vanishing),
while for vertices where only one layer has a one-dimensional decomposition, 
this yields new effective plaquette terms $\rho_p$ in the other layer by projecting
the original plaquette terms $\Pi_p$ onto that one-dimensional subspace; thus,
the problem of checking whether $\Omega(\vec\alpha,\vec\beta)$ is non-zero
reduces to computing the overlap of the new effective plaquette terms
$\rho_p$.
 Formally, this reads
\begin{align}
        \nonumber
\Omega(\vec\alpha,\vec\beta) &= 
\tr\Bigg[\,
\Big(\!\!\bigotimes_{p\in \mc P_B} \Pi_p^{\vec\alpha}\,\Big)
\Big(\!\!\bigotimes_{p\in \mc P_W} \Pi_p^{\vec\beta}\,\Big)
\,\Bigg]
\\[2ex]
&\stackrel{\llap{\mbox{\scriptsize(A)}}}{=}
\tr\Bigg[\,
\underbrace{
\Big(\!\prod_{w\in \mc F_W}\!\!\bar\pi^w_{\beta_w}
\bigotimes_{p\in \mc P_B}\!\! \Pi_p^{\vec\alpha}
\prod_{w\in \mc F_W}\!\!\bar\pi^w_{\beta_w}\Big)}_{(*)}\;
\Big(\!\!\!\prod_{v\in \mc F_B\backslash \mc F_W}
    \!\!\!\!\!\!\!\pi^v_{\alpha_v}
\bigotimes_{p\in \mc P_W}\!\! \Pi_p^{\vec\beta}
\prod_{v\in \mc F_B\backslash\mc F_W}
    \!\!\!\!\!\!\!\pi^v_{\alpha_v} \Big)\;
\,\Bigg]\nonumber\\[2ex]
&\stackrel{\llap{\mbox{\scriptsize(B)}}}{=}
\prod_{v\in\mc F_B\cap \mc F_W}\!\!\!\!\!
    \tr\big[\pi^v_{\alpha_v}\bar\pi^v_{\beta_v}\big]\ 
    \times\ 
    \tr\Bigg[
    \,\,\Big(\!\!\bigotimes_{p\in \mc P_B} \rho_p\,\Big)
    \Big(\!\!\bigotimes_{p\in \mc P_W} \rho_p\,\Big)
    \Bigg]\ .
    \label{eq:afterb}
\end{align}
\vspace*{1ex}
Here, we have used in step (A) that for all $v\in \mc F_W$,
\[
\bigotimes_{p\in \mc P_W} \Pi_p^{\vec\beta}\,
=
\bar\pi^v_{\beta_v}\,
\Big(\!\!\bigotimes_{p\in \mc P_W} \Pi_p^{\vec\beta}\,\Big)
\;\bar\pi^v_{\beta_v}
\]
and anologously for the black layer; in step (B), we have defined 
new effective plaquette terms $\rho_p$ by virtue of
\[
\bigotimes_{p\in \mc P_B}\!\! \rho_p
=
\tr_{\mc F_B\cup \mc F_W}\Bigg[
\Big(\!\prod_{w\in \mc F_W\backslash \mc F_B}\!\!\bar\pi^w_{\beta_w}
\bigotimes_{p\in \mc P_B}\!\! \Pi_p^{\vec\alpha}
\prod_{w\in \mc F_W\backslash \mc F_B}\!\!\bar\pi^w_{\beta_w}\Big)
\Bigg]\ ,
\]
and correspondingly for the white plaquettes; the first factor in
(\ref{eq:afterb}) takes
care of the terms in $\mc F_B\cap \mc F_W$ in $(*)$. Note that the
$\rho_p$ are now only supported on those vertices not in $\mc F_B\cup \mc
F_W$, as those have been traced out.

The task of checking whether $\Omega(\vec\alpha,\vec\beta)\ne 0$ has thus
been reduced to checking this for (\ref{eq:afterb}): For the first term,
this can be clearly done efficiently, and so, it remains to prove that the
overlap
\begin{equation}
        \label{eq:theta}
\Theta:= \tr\Bigg[
    \,\,\Big(\!\!\bigotimes_{p\in \mc P_B} \rho_p\,\Big)
    \Big(\!\!\bigotimes_{p\in \mc P_W} \rho_p\,\Big)
    \Bigg]
\end{equation}
can be computed efficiently. Note that the $\rho_p$ are now supported on
plaquettes of a square lattice with vertices missing. Moreover, while the
$\rho_p$ do no longer commute, in each layer at most one $\rho_p$ acts
non-trivially on each vertex; we will make use of this fact repeatedly in
the following.

\begin{figure}
\begin{center}
\includegraphics[width=6cm]{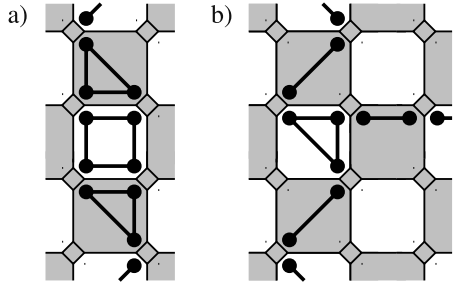}
\parbox{0.9\textwidth}{
\caption{\label{fig:1d-2d-chains}
Computing the overlap Eq.~(\ref{eq:theta}) for qubits.  The diamonds at
the vertices of the square lattice denote the qubits.  The connected black
dots mark
qubits on which the $\rho_p$ act non-trivially on plaquette $p$. If
the $\rho_p$ of two plaquettes act non-trivially on the same qubit, we say
that they ``overlap''; note that this cannot happen for diagonally
adjacent plaquettes.  Overlapping $\rho_p$'s form structures which we
need to contract to compute the overlap Eq.~(\ref{eq:theta}).
\textbf{a)} Patterns forming one-dimensional chains can be contracted
efficiently, as the size of the boundary stays constant for any contiguous
region. \textbf{b)} Branching structures  do in general not allow for
efficient contraction. However, we show that such structures cannot occur,
by proving that for any plaquette $p$, $\rho_p$ can overlap non-trivially
at most with two adjacent $\rho_{p'}$'s.
}}
\end{center}
\end{figure}

The situation encountered in computing the overlap $\Theta$ is depicted in
Fig.~\ref{fig:1d-2d-chains}.  Here, the dots in each plaquette denote the
vertices on which $\rho_p$ acts non-trivially (the lines just connect the
vertices involved in $\rho_p$). If the $\rho_p$ on adjacent plaquettes act
non-trivially on the same qubit (we will say they ``overlap''), they form
connected structures which we need to contract in order to evaluate
$\Theta$.  For one-dimensional structures as the one on in
Fig.~\ref{fig:1d-2d-chains}a, this contraction can be carried out
efficiently: One starts from one pla\-quette and proceeds along one
direction of the one-dimensional chain, always tracing out the degrees of
freedom on the inside. This way, at every point in the computation only
the state at the boundary (which has fixed size) needs to be stored, and
thus, the contraction can be carried out efficiently.  On the other hand,
branching structures like the one in Fig.~\ref{fig:1d-2d-chains}b can in
general not be contracted efficiently, since the size of the boundary is a
priori not bounded; in fact, e.g.\ quantum circuits, and even postselected
quantum circuits, can be encoded this way, making such contractions in
general a computationally hard task. 

However, as we will show in the following, the structures formed by the
$\rho_p$ in $\Theta$, Eq.~(\ref{eq:theta}), will always be
one-dimensional, and thus $\Theta$ can be computed efficiently.  To this
end, we will consider the state $\rho_C$ on a plaquette $C$ (the
``central'' plaquette), and show that it can overlap non-trivially with
the states $\rho_p$ of at most two of the adjacent plaquettes, thus ruling
out branching structures as the one on the right of
Fig.~\ref{fig:1d-2d-chains}.  We will make intensive use of the fact that
in each layer, at most one plaquette term $\rho_p$ can act non-trivially
on any given vertex; in the graphical notation of
Fig.~\ref{fig:1d-2d-chains}, we will highlight this fact by placing a
cross opposite of any dot: 
\begin{equation}
        \label{eq:dot-cross}
\raisebox{-0.35cm}{\includegraphics[height=1cm]{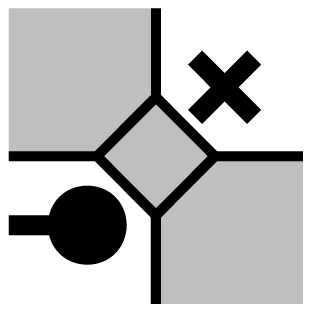}}
\end{equation}
Here, the dot indicates that the corresponding $\rho_p$ acts non-trivially
on a vertex qubit (the diamond in the center), while the cross indicates that
the corresponding $\rho_p$ does act trivially. Note that this in
particular implies that $\rho_C$ can at most overlap non-trivially with the
four horizontally and vertically adjacent plaquettes from the other layer,
but not with diagonally adjacent plaquettes.

The simplest case is when the state $\rho_C$ on the central plaquette $C$
involves only two vertices non-trivially, for instance 
\[
\includegraphics[height=2.5cm]{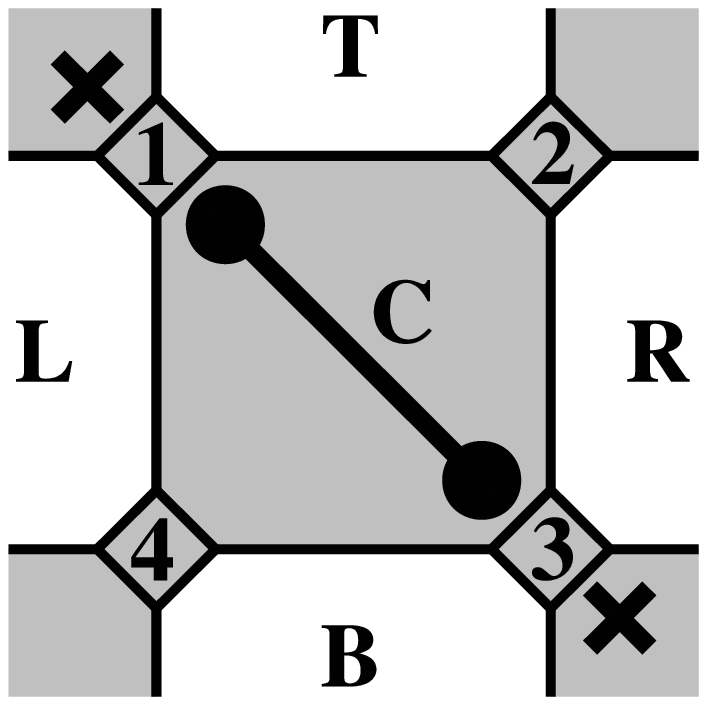}
\]
Now, both qubits $1$ and $3$ can be acted upon non-trivially by at most one
white plaquette -- the other has to be empty, following the rule
(\ref{eq:dot-cross}) that opposite of any dot there has to be a cross;
this way, only one-dimensional structures can arise:
\[
\includegraphics[height=2.5cm]{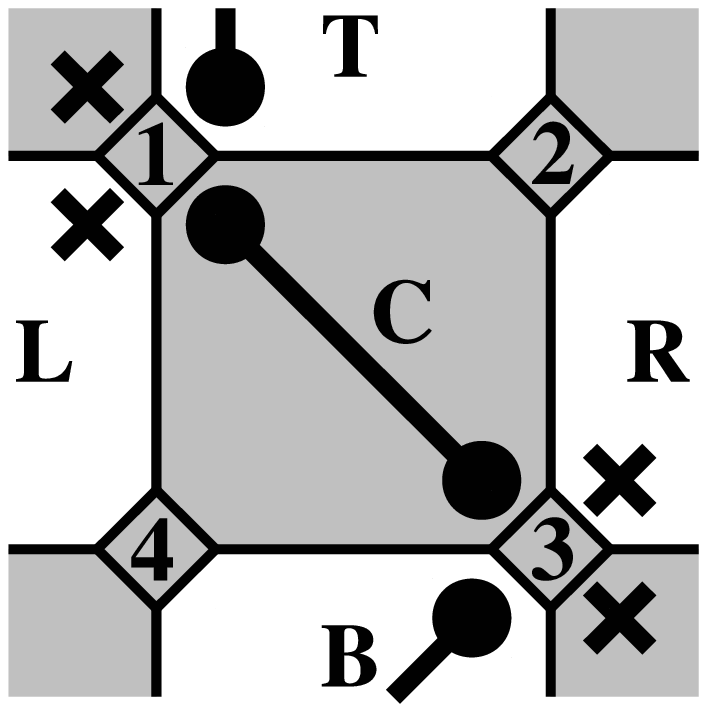}
\]
This clearly holds for any possible $\rho_C$ which acts non-trivially on
only two qubits, and for any configuration of the adjacent plaquettes; it
follows that only one-dimensional structures can emerge this way.

In order to understand the cases where $\rho_C$ acts non-trivially on
three or four qubits, let us first analyze the following situation:
\[
\includegraphics[height=1.25cm]{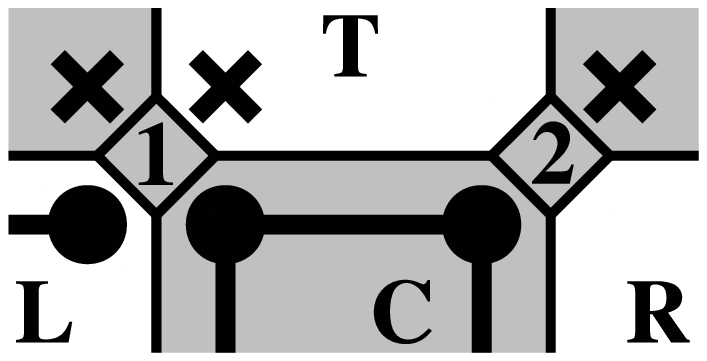}
\]
Here, $\rho_C$ acts non-trivially at least on qubits $1$ and $2$, and
$\rho_L$ acts non-trivially on qubit $1$ which implies that $\rho_T$
acts trivially on it.  In the following, we will show that this implies
that $\rho_T$ also has to act trivially on qubit $2$. We will prove this
by contradiction, so assume that $\rho_T$ acts non-trivially on qubit $2$.
Since $\rho_T$ is obtained from the original projector $\Pi_T$ on that
plaquette by a partial projection on some of the other vertices,
this implies that $\Pi_T$ acts non-trivially on qubit $2$ (where it spans
the whole $\cs$--algebra, since we have traced out all vertices where this
was not the case).  On the other hand, $\rho_L$ and thus $\Pi_L$ acts
non-trivially on qubit $1$, and thus spans the whole $\cs$--algebra on it.
Since $\Pi_L$ and $\Pi_T$ commute, this means that $\Pi_T$ acts trivially
on qubit $1$; that is, $\Pi_T$ and $\Pi_C$ need to commute on qubit $2$
alone: However, since $\Pi_T$ spans the whole $\cs$--algebra on qubit $2$, 
this would imply that $\Pi_C$ and thus $\rho_C$ has to act
trivially on qubit $2$, giving a contradiction.
Thus, we have the following ``Lemma'':
\begin{equation}
        \label{eq:central-lemma-qubits}
\raisebox{-0.5cm}{\includegraphics[height=1.25cm]{central-step-left}}
\qquad{\bm\Longrightarrow}\qquad
\raisebox{-0.5cm}{\includegraphics[height=1.25cm]{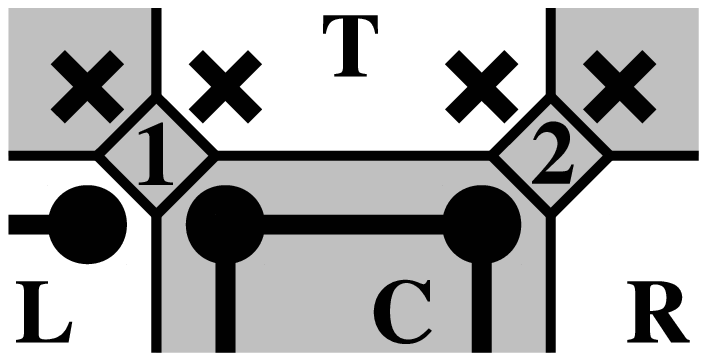}}
\end{equation}

Let us now consider the case where the state on the central plaquette
involves all four qubits non-trivially, and let us start by assuming
w.l.o.g.\ that
$\rho_L$ acts non-trivially on qubit~$1$:
\[
\includegraphics[height=2.5cm]{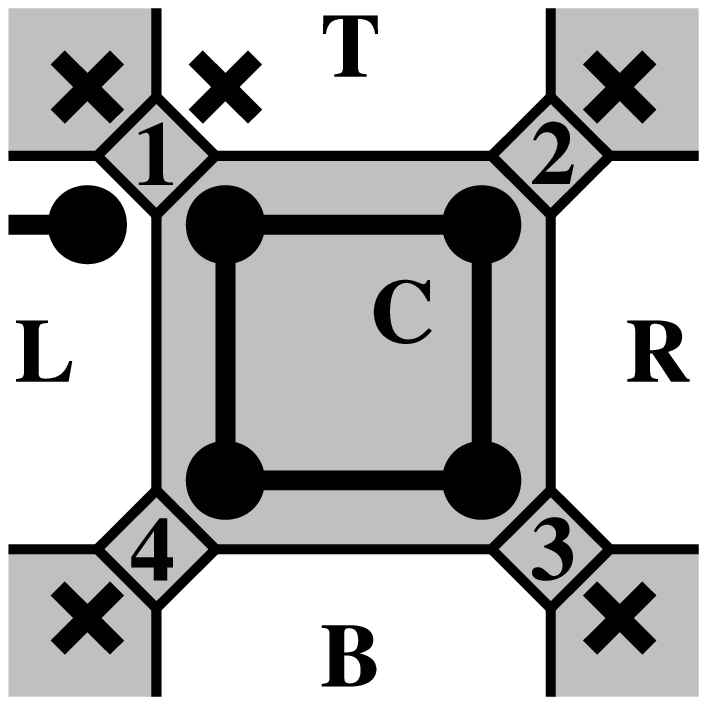}
\]
This implies that qubit $1$ is not acted upon by $\rho_T$. Using
Eq.~(\ref{eq:central-lemma-qubits}), we infer that $\rho_T$ cannot involve
qubit $2$ either,
\[
\includegraphics[height=2.5cm]{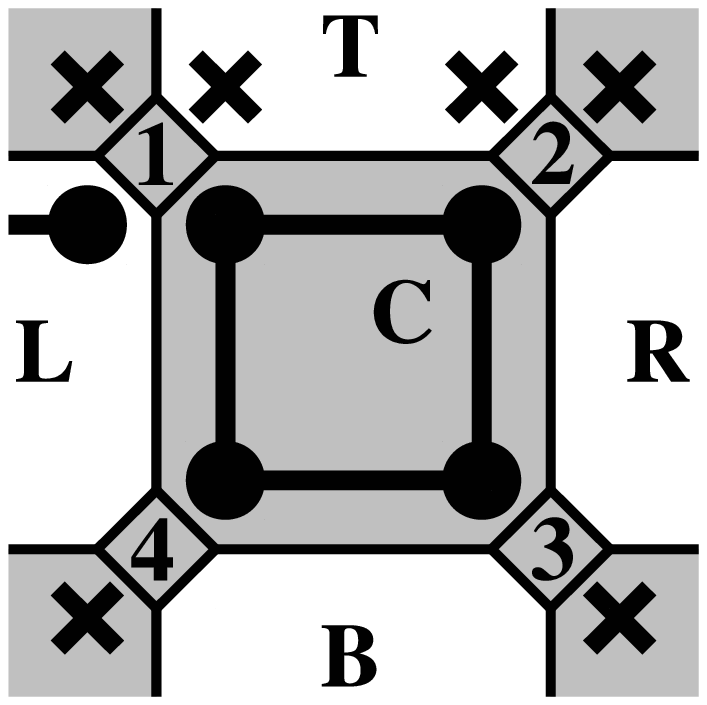}
\]
Eq.~(\ref{eq:central-lemma-qubits}) also shows that
$\rho_B$ has to act trivially
on qubit $4$ -- otherwise, $\rho_L$ would act trivially on qubit $4$ and
thus qubit $1$, which it doesn't: 
\[
\includegraphics[height=2.5cm]{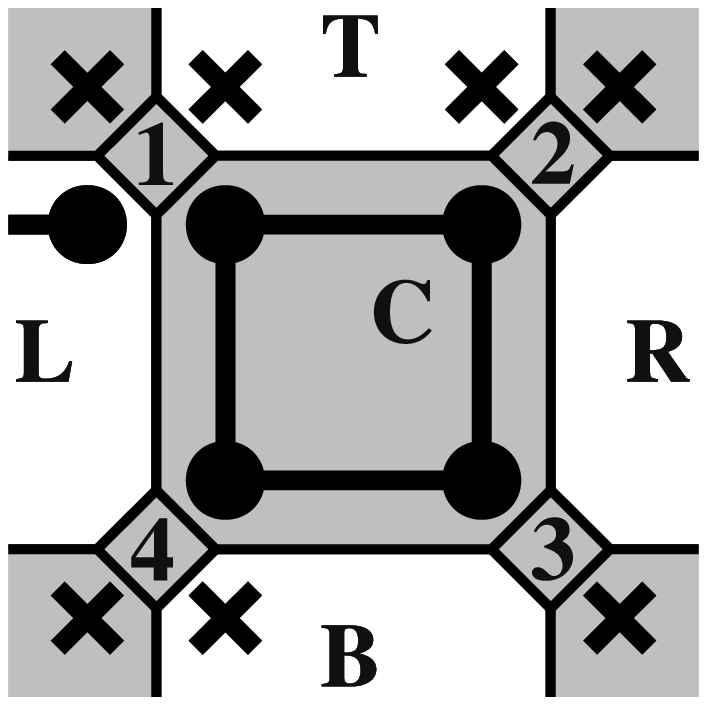}
\]
In order to obtain a branching structure, both $\rho_B$ and $\rho_R$ need
to act non-trivially on some of the qubits. However, if $\rho_B$
acts non-trivially on qubit $3$, Eq.~(\ref{eq:central-lemma-qubits})
implies that $\rho_R$ has to act trivially on qubits $2$ and $3$:
\[
\includegraphics[height=2.5cm]{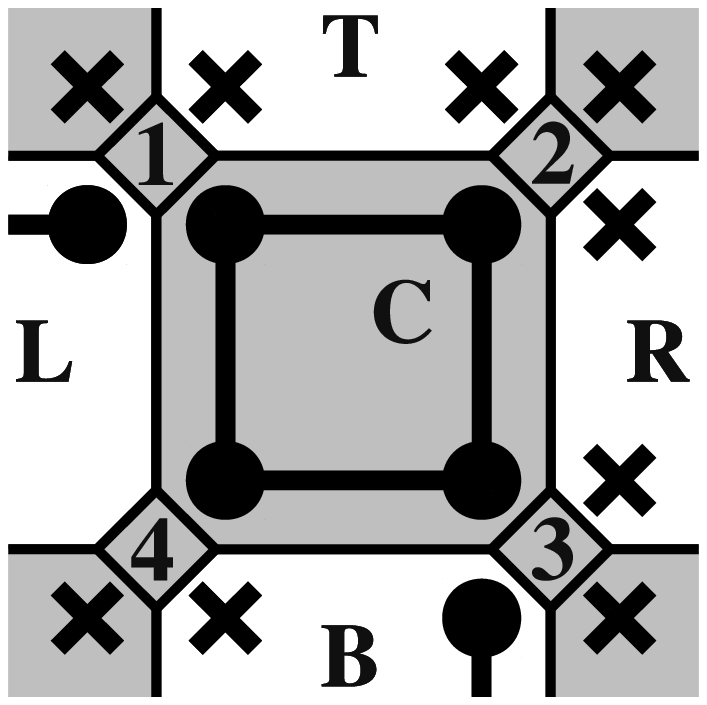}
\]
This shows that for $\rho_C$ acting non-trivially on all four qubits, the
central plaquette can only couple to at most two adjacent plaquettes,
forming one-dimensional structures.

It remains to study what happens in the case of tripartite entanglement on
the central plaquette. We start from the following configuration:
\[
\includegraphics[height=2.5cm]{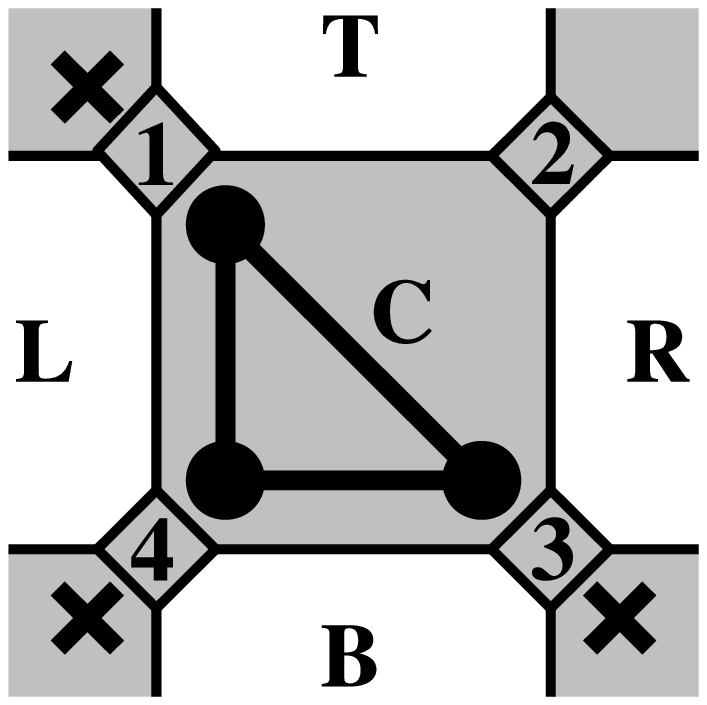}
\]
(Note that we don't make any assumptions on how the Hilbert space of qubit
$2$ decomposes.)  Clearly, in order to obtain a branching structure, either
$\rho_T$ has to act non-trivially on qubit $1$, or $\rho_R$ has to act
non-trivially on qubit $3$.
We consider w.l.o.g.\ the first
possibility and infer from Eq.~(\ref{eq:central-lemma-qubits}) that
$\rho_L$ has to act trivially on both qubits $1$ and $4$:
\[
\includegraphics[height=2.5cm]{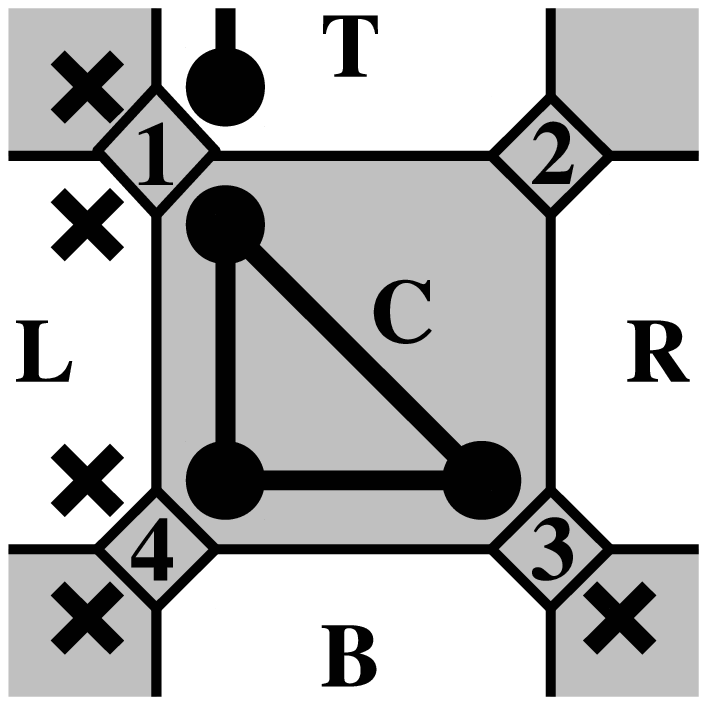}
\]
In order to obtain a branching structure,  we now have to get both $\rho_R$
and $\rho_B$ involved. However, if $\rho_R$ acts non-trivially on qubit
$3$, Eq.~(\ref{eq:central-lemma-qubits}) implies that $\rho_B$
has to act trivially on both qubits $3$ and $4$, and thus, the structure
formed around $\rho_C$ will again be one-dimensional.

Together, this shows that the overlap $\Theta$, Eq.~(\ref{eq:theta}),
decays into one-dimen\-sional structures for which the overlap can be
computed efficiently.  In turn, this implies that for given $\vec\alpha$
and $\vec\beta$, $\Omega(\vec\alpha,\vec\beta)$ can be computed
efficiently, and thus, the commuting Hamiltonian problem on a square
lattice of qubits with plaquette interactions is in \textsf{NP}.

\subsection{Finite accuracy}

In the preceding proof, we have assumed infinite accuracy, but as we will
now show, our argument still applies if we compute with finite accuracy.
To this end, let $N$ denote the number of qubits in the system; we will
need to show that the computation time scales as $\poly(N)$.  We assume
that the Hamiltonian terms are given exactly and can be represented with
$\poly(N)$ digits.  First, note that the trace in
Eq.~(\ref{eq:commham-trace-formulation}), which equals the sum in
Eq.~(\ref{eq:sum_alphabeta_omega}), evaluates to an integer, and thus,
there exists at least one pair $(\vec\alpha,\vec\beta)$ such that
$\Omega(\vec\alpha,\vec\beta)\ge 2^{-2N}$. If we request this particular
$(\vec\alpha,\vec\beta)$ as a certificate, it is sufficient if we can
determine $\Omega(\vec\alpha,\vec\beta)$ to $2N+1=\poly(N)$ digits. (This
is crucial, since there can be $(\vec\alpha,\vec\beta)$ for
which $\Omega(\vec\alpha,\vec\beta)$ is arbitrarily small.)
$\Omega(\vec\alpha,\vec\beta)$ is obtained from contracting a polynomial
number of terms which are either $\Pi_p$ (which are known exactly) or
$\pi^v_{\alpha_v}$ and $\bar\pi^v_{\beta_v}$, and the latter can be
determined to $\poly(N)$ accuracy from the
$\cs$--decomposition~(\ref{eq:cstardec}), which is the solution to a
(fixed-size) eigenvalue problem. It follows that
$\Omega(\vec\alpha,\vec\beta)$ can be computed to the required $\poly(N)$
accuracy in $\poly(N)$ time, and our proof still applies.

\section{Summary}

We have studied the \textsc{commuting hamiltonian} problem on a square
lattice of qubits with plaquette-wise interaction and shown that the
problem is \textsf{NP}--complete.  Differently speaking, we have shown
that there exists a classical certificate for the fact that the ground
state of the system minimizes each term locally which can be checked
efficiently on a classical computer. The central idea for our proof has
been to split the system into two layers in each of which the commuting
terms overlap on individual sites, and to argue that the existence of a
state minimizing all local terms is equivalent to the existence of a pair
of ground states for the two layers with non-zero overlap. Each layer
could be decomposed using the $\cs$--algebraic techniques introduced to
the problem in~\cite{bravyi:comm-ham}, allowing to find an efficient
description of its ground state subspace. Finally, we showed that the
overlap of ground states of two layers can be computed efficiently, by
showing that it gives rise to of one-dimensional structures only.  A
somewhat surprising feature of our approach is that while it certifies the
existence of a ground state, it cannot (to our knowledge) be used to
devise a way how to prepare the ground state; in fact, due to the
possibility of having topological order in such systems, any circuit
preparing their ground states would need to have at least logarithmic
depth, or linear depth if it was
local~\cite{bravyi:lr-bound-topo-order}.

Our method does, in principle, also apply beyond qubits: We can still
split the system into two layers, decompose both of them into direct
sum slices $\vec\alpha$ and $\vec\beta$, and ask the prover to provide
labels $\vec\alpha$ and $\vec\beta$ with non-zero overlap
$\Omega(\vec\alpha,\vec\beta)$.  While we cannot make sure any more that
$\Omega(\vec\alpha,\vec\beta)$ can be computed efficiently, we can always
ask the prover to also provide us with an instruction on how to
efficiently contract the states, in case there is a way to do so, e.g.\
by providing the optimal contraction order.  In particular, this
applies to the case where the decomposition in the direct sum gives
one-dimensional spaces, such as in Kitaev's toric code or quantum double
models; as well as to cases where the $\rho_p$ are separable states. Our
idea also applies to any other graph which can be split into two layers in
such a way that the $\cs$--technique of~\cite{bravyi:comm-ham} can be
applied to each of them, and in fact to any type of Hamiltonian which is
composed of two layers with eigenbases for the zero-energy subspace whose
overlap can be computed efficiently, such as for product bases. Note that
on the other hand, a decomposition into three layers cannot be used for
our purposes, since for three positive operators $A$, $B$, and $C$, 
$\tr[ABC]$ can have both real and imaginary parts of either sign, so
that Eq.~(\ref{eq:sum_alphabeta_omega}) is no longer equivalent to
Eq.~(\ref{eq:def-omega}) being non-zero (as the $\Pi_p^{\vec\alpha}$ and 
$\Pi_p^{\vec\beta}$ do not commute any more).

An interesting open question relating to the present approach to the
problem is whether it can be generalized beyond qubits. For four-level
systems and beyond, this is likely not the case, since the local Hilbert
space can decompose into two qubits, and thus operators commuting on a
single spin can both act non-trivially on it, i.e.,
Eq.~(\ref{eq:dot-cross}) does not hold any more.  On the other hand, for
qutrits this is not the case once we have fixed a slice in the direct sum;
yet, it is not clear how to establish a version of
Eq.~(\ref{eq:central-lemma-qubits}).  In particular, the non-trivial
projections $\pi_{\alpha_v}^v$ can now have both rank $1$ and $2$, and in
the latter case we cannot simply trace out the corresponding degree of
freedom; it is however not clear that this does rule out an analogue to
Eq.~(\ref{eq:central-lemma-qubits}).

\section*{Acknowledgements}
We acknowledge helpful conversations with
Dorit Aharonov, 
Sergey Bravyi,
Lior Eldar,
Tobias Osborne,
and
Volkher Scholz.
This work has been supported by the 
Gordon and Betty Moore Foundation
through Caltech's Center for the Physics of Information,
the NSF Grant \mbox{No.~PHY-0803371}, and the ARO Grant
No.~W911NF-09-1-0442.

\end{document}